\documentclass[11pt,a4paper]{article}
\usepackage{hepunits}
\usepackage{slashed}
\usepackage{multirow}
\usepackage{jheppub}

\newcommand{\neu}[1]{\ensuremath{\tilde{\chi}_{#1}^0}\xspace}

\newcommand{\met}{\ensuremath{E^{\text{miss}}_T}\xspace}

\newcommand{\mgrv}{\ensuremath{m_{3/2}}\xspace}
\newcommand{\ftno}{\texttt{Fittino}\xspace}
\newcommand{\msg}{\ensuremath{m_{\tilde{g}}}\xspace}
\newcommand{\msq}{\ensuremath{m_{\tilde{q}}}\xspace}
\newcommand{\tev}[1]{\unit{#1}{\TeV}}
\newcommand{\gev}[1]{\unit{#1}{\GeV}}
\newcommand{\ifb}[1]{\unit{#1}{\invfb}}

\newcommand{\sell}{\ensuremath{\tilde{\ell}}\xspace}
\newcommand{\chpm}[1]{\ensuremath{\tilde{\chi}_{#1}^{\pm}}\xspace}
\newcommand{\sq}{\ensuremath{\tilde{q}}\xspace}
\newcommand{\stau}{\ensuremath{\tilde{\tau}}\xspace}
\newcommand{\sg}{\ensuremath{\tilde{g}}\xspace}
\newcommand{\sd}{\ensuremath{\tilde{d}}\xspace}
\newcommand{\su}{\ensuremath{\tilde{u}}\xspace}
\newcommand{\st}{\ensuremath{\tilde{t}}\xspace}

\newcommand{\comment}[1]{}

\title{Using rates to measure mixed modulus-anomaly mediated supersymmetry breaking
  at the LHC}
\author[a]{J. A. Conley,}
\author[a]{H. K. Dreiner,}
\author[b]{L. Glaser,}
\author[b]{M. Kr\"amer,}
\author[a,b]{and J. Tattersall}
\affiliation[a]{Physikalisches Institut der Universit\"at Bonn\\
Nussallee 12, D-53115 Bonn, Germany}
\affiliation[b]{Institute for Theoretical Particle Physics and Cosmology\\
RWTH Aachen, D-52056 Aachen, Germany}
\emailAdd{conley@th.physik.uni-bonn.de}
\emailAdd{dreiner@th.physik.uni-bonn.de}
\emailAdd{Lisa.Glaser@physik.rwth-aachen.de}
\emailAdd{mkraemer@physik.rwth-aachen.de}
\emailAdd{jamie@th.physik.uni-bonn.de}

\abstract{If SUSY is discovered at the LHC, the task will immediately turn to
determining the model of SUSY breaking. Here, we employ a Mixed
Modulus-Anomaly Mediated SUSY Breaking (MMAMSB) model with very
similar LHC phenomenology to the more 
conventionally studied Constrained Minimal SUSY Model (CMSSM) and minimal Anomaly Mediated SUSY Breaking 
(mAMSB) models. We then study whether the models can be distinguished and measured. If we only fit to the various 
mass edges and mass end-points from cascade decay chains that are normally studied, a unique determination and 
measurement of the model is problematic without substantial amounts of LHC data. However, if event rate 
information is included, we can quickly distinguish and measure the correct SUSY model and exclude alternatives.}

\begin{document}
\begin{flushright}BONN-TH-2011-14\\TTK-11-46\end{flushright}
\maketitle

\section{Introduction}



The experiments at the Large Hadron Collider (LHC) are currently
running and have begun to explore physics at the Terascale.
In particular, the LHC has already probed large regions of the
parameter space of supersymmetry (SUSY), one of the most
well motivated and thoroughly studied approaches to solving the
hierarchy problem \cite{Aad:2011hh,daCosta:2011qk,Aad:2011ks,
  Aad:2011xm,ATLAS-CONF-2011-090,ATLAS-CONF-2011-098,
  jetsPlusMissingEPS,CMS-PAS-SUS-11-010,Chatrchyan:2011wc,
  Chatrchyan:2011ah,Chatrchyan:2011bj,Chatrchyan:2011ek,
  Chatrchyan:2011qs}.

Generically, the dominant signature for R-parity-conserving SUSY at
the LHC is events with high transverse momentum ($p_T$) jets (and
maybe leptons) accompanied by large amounts of missing transverse
energy (\met), originating from the pair production of squarks and
gluinos which then cascade decay eventually to the lightest neutralino
(LSP). Indeed, the strongest limits to date on the
supersymmetric parameter space come from searches for events with
multiple hard jets plus missing energy and zero or one lepton
\cite{ATLAS-CONF-2011-090,jetsPlusMissingEPS,Chatrchyan:2011qs,
  Chatrchyan:2011ek}.

If new physics consistent with SUSY is discovered at the LHC, these
cascade decay events may also be instrumental in determining the
Lagrangian parameters of the underlying model. Various kinematic edges can be
constructed from the jets, leptons, and missing energy. The
measurement of these edges then constrains the SUSY mass
spectrum \cite{Hinchliffe:1996iu,Allanach:2000kt,Gjelsten:2004ki,
  Weiglein:2004hn}. Combined with additional measurements of sparticle 
decay branching ratios, these measurements can be used as input into a
global fit for the parameters of a given SUSY model.

Several groups have performed such fits for various supersymmetric
parameter spaces \cite{Lester:2005je,Bechtle:2009ty,Adam:2010uz,Bertone:2011nj,
  Buchmueller:2011ki,Allanach:2011ya}. In a
sufficiently low-dimensional parameter space such as the constrained
minimal supersymmetric standard model (CMSSM) and with sufficient luminosity, the LHC
measurements are powerful enough to enable a precise determination of
the best-fit SUSY parameters. With low luminosity and/or in a
parameter space with a greater number of degrees of freedom, however,
the endpoint and branching ratio measurements alone are not always
sufficient to ensure a stable fit and a precise parameter
determination. 

In Ref.~\cite{Lester:2005je}, it was demonstrated that the inclusion 
of a cross section measurement---in particular the 
rate of high-missing-$p_T$ events---can drastically improve the precision
of a fit to a SUSY parameter space with non-universal gaugino
masses. The authors of Ref.~\cite{Lester:2005je} use a standard 
leading-order Monte Carlo simulation and a fast detector simulation
to compute the rate at each point in parameter space. While this 
method is straightforward and general, there are some technical challenges
to this approach that have so far made it impractical to include
rates in most such fits.

The first issue is the computational cost of a Monte Carlo simulation.
In Ref.~\cite{Lester:2005je}, this is addressed by only generating
a small number (1000) of events per parameter space point, and by
implementing a parallelized version of the Monte Carlo generator 
\texttt{Herwig} \cite{Corcella:2000bw} on multiple processors.
Even so, the computation of rates is extremely time-consuming
and it is only feasible to scan a relatively small number of parameter
space points when doing the fit. 
Because of the limitation on the number of events that can be
generated per point in parameter space, there are also large statistical
fluctuations in the computed rate. These fluctuations can spoil the 
stability of the fit. Finally, while the state-of-the-art calculations
of SUSY production cross sections at the LHC include many higher-order
corrections, the Monte Carlo generators only use leading-order
cross sections.

Recently, a technique to include rate information in fits and overcome 
the issues listed above was presented in Ref.~\cite{Dreiner:2010gv},
and was implemented in the fitting code \texttt{Fittino} \cite{Bechtle:2004pc}.
In this implementation, the cross section for sparticle production
is first interpolated from a grid in the space of sparticle masses
that was pre-computed using the
program \texttt{Prospino} \cite{Beenakker:1996ed}.
The remaining factor in
the event rate, the acceptance---the fraction of sparticle production
events whose decay products pass the kinematical cuts---is computed
using a novel semi-analytical method. Because computing the rate in this way 
is extremely fast and sufficiently accurate, it allows event rates to be
used on the same footing as the usual edge measurements in fits to
LHC data. We use this technique as implemented in \texttt{Fittino}
in the fits performed here and find that the rates are often a crucial ingredient
in obtaining convergent and precise measurements of SUSY parameters.

Of course, we do not a priori know the mechanism of supersymmetry
breaking, and thus which SUSY parameter space is the appropriate one
in which to run such a fit. Most fits have concentrated on the CMSSM
parameter space, because it involves a small number of parameters and
because it has been extremely well studied.  One could of course ask,
would LHC measurements provide sufficient information to select one
model of SUSY breaking over another? In other words, would a fit
performed in the ``wrong'' parameter space yield a significantly worse
goodness-of-fit than one performed in the ``right'' parameter space?
This question was recently investigated \cite{Allanach:2011ya} for a
number of popular SUSY breaking models: the CMSSM, which is a version of
gravity-mediated SUSY breaking; minimal gauge-mediated SUSY breaking
(mGMSB); minimal anomaly mediation (mAMSB) and large-volume string
compactification models. Assuming a signal with \ifb{1} of LHC data at $\sqrt{s}
= 14$~TeV it was found that the CMSSM scenario can be
distinguished, using the LHC endpoint measurements only, from all
other scenarios considered except for the mGMSB, which can fit the
data considered equally well.

One SUSY breaking scenario that until now has not been studied in the
context of a global fit is mirage mediation, also known as mixed
modulus-anomaly mediated SUSY breaking (MMAMSB). This model is
attractive theoretically because it can be derived from a concrete string
compactification scenario due to Kachru, Kallosh, Linde and Trivedi
(KKLT) which provides mechanisms for breaking SUSY, stabilizing
unwanted moduli, and ensuring a positive cosmological constant
\cite{Kachru:2003aw}. In fact in Ref.~\cite{Choi:2007ka} it is argued
that the MMAMSB breaking pattern is a generic feature of a certain
class of mechanisms that stabilize string moduli, of which the KKLT
construction is only one example. 

A couple of papers have analyzed the soft SUSY
breaking terms that arise from specific realizations of this scenario,
pointing out that the SUSY-breaking contributions from gravity
(i.e.\ modulus) mediation are comparable to those coming from anomaly
mediation \cite{Choi:2004sx,Choi:2005ge}. A parameter of the model,
$\alpha$, interpolates between pure anomaly mediation and pure modulus
mediation. For vanishing $\alpha$, the model suffers from the
tachyonic slepton problem of pure AMSB, whereas for intermediate or
large values of $\alpha$, this problem is alleviated by the gravity-mediated
contributions.

The phenomenology of MMAMSB has been studied in several papers 
\cite{Choi:2005uz,Falkowski:2005ck,Endo:2005uy,Baer:2006id,Baer:2006tb,
  Cho:2007fg,Baer:2007eh,Altunkaynak:2010xe}, where it has been shown
that there are good prospects for discovering this type of SUSY breaking
scenario at the LHC and/or at direct dark matter detection experiments. At least
for certain choices of the model parameters, it should also be
possible to measure them and identify the gaugino mass
pattern distinctive to this model.

In this paper, we aim to quantify the extent to which the
MMAMSB scenario can be explored at the LHC by performing a global fit
to prospective LHC measurements in the MMAMSB parameter space. This way, we can
evaluate the accuracy and precision of the measurement of the model
parameters. By attempting to fit different SUSY-breaking scenarios,
the CMSSM and mAMSB, to the MMAMSB model, we can also evaluate whether LHC
measurements are sufficient to distinguish MMAMSB from other
SUSY-breaking scenarios. We show that the inclusion of rates as
inputs to the fit is necessary to enable accurate parameter
measurements with relatively low luminosity, and that rates are especially
crucial in distinguishing MMAMSB from the CMSSM and mAMSB.

The layout of this paper is as follows. We begin by giving a brief introduction to the MMAMSB model and, in particular, the phenomenology of the gaugino sector. In Sec.~\ref{sec:Model_point}, we discuss choosing a particular benchmark point in the model that satisfies all of the current experimental and observational constraints. In Sec.~\ref{sec:Fit_proc}, we discuss how we fit the model with hypothetical LHC data and introduce the different observables that we use. Finally, in Sec.~\ref{sec:Fit_res} we present the results of fitting the MMAMSB to observables derived from our benchmark point. We also try to fit other SUSY breaking models (the CMSSM and mAMSB) to our benchmark to see how quickly these models can be excluded and which observables are effective at performing this task. We conclude with a discussion in Sec.~\ref{sec:Disc}.

\section{Mixed moduli-anomaly mediated supersymmetry breaking}



In this section we briefly introduce the MMAMSB scenario. We
adopt the notation of Ref.~\cite{Baer:2006id}. 
As mentioned in the introduction, the MMAMSB can be derived from the
KKLT string 
construction \cite{Kachru:2003aw}, or indeed from a class of string
models in which moduli stabilization and SUSY breaking are
accomplished dynamically \cite{Choi:2007ka}. In these models, the soft
SUSY breaking terms receive contributions from both modulus and
anomaly mediation. The parameter $\alpha$ interpolates between the
pure modulus and pure anomaly extremes.

The soft terms also depend on
the sets of parameters $n_i$ and $\ell_a$, which correspond to powers in the K\"ahler
potential for matter fields and the gauge kinetic functions respectively.
In the KKLT construction, $\ell_a$ can take on the value 0 or
1 depending on whether the corresponding gauge field is localized on a $D3$ or
$D7$ brane, whereas $n_i$ can be 0, 1, or $1/2$ for a matter field
localized on a $D7$ brane, a $D3$ brane, or a brane intersection. 

Here we concentrate on the scenario where $\ell_a=1$ for all
gauge fields and where $n_i=n$ for all matter fields (including the
Higgs fields). Having a common value for $n_i$ among the sfermions is
motivated by flavor constraints. Having a different $n_i$ for the
sfermions and the Higgses is possible, but for simplicity we assume a
common value here. With these assumptions, the soft SUSY breaking
parameters---the gaugino mass parameters, trilinear couplings, and
sfermion mass parameters, respectively---at the GUT scale are given by
\begin{eqnarray} 
  M_a &=& \frac{m_{3/2}}{16\pi^2}[\alpha + b_ag^2_a], \label{eq:Gauge_MM} \\
  A_{ijk} &=& \frac{m_{3/2}}{16\pi^2}[3(n-1)\alpha + 
  (\gamma_i + \gamma_j + \gamma_k)], \label{eq:Trilinear_MM} \\
  m^2_i &=& \left(\frac{m_{3/2}}{16\pi^2}\right)^2[(1-n)\alpha^2 + 
  4\alpha\xi_i - \dot{\gamma}_i]. \label{eq:Scalar_MM}
\end{eqnarray}
Here, $m_{3/2}$ is the gravitino
mass, $g_a$ is the gauge coupling and $b_a=(\frac{33}{5},1,-3)$ the
1-loop $\beta$ function coefficient for the gauge group $a$. The anomalous
dimensions $\gamma_i$ and their logarithmic derivatives $\dot{\gamma}_i$ as
well as the mixed anomaly-modulus contributions $\xi_i$ are given in
Appendix A of Ref.~\cite{Falkowski:2005ck}.

In Eq.~(\ref{eq:Gauge_MM}) the term proportional to $\alpha$ is the
universal gravity mediation contribution, whereas the second term is
the anomaly mediation contribution. The ratio of gauge couplings at
the TeV scale should be a combination of the expected ratio from
gravity mediation, 
\begin{equation}
  M_1:M_2:M_3 \simeq 1:2:6\;,
\end{equation}
and the expected ratio from anomaly mediation,
\begin{equation}
  M_1:M_2:M_3 \simeq 3.3:1:9\;.
\end{equation}
For the MMAMSB,
it is\footnote{In this paper we follow the notation of 
  Ref.~\cite{Baer:2006id} by Baer et al., in particular
  $\alpha=\alpha_{\text{Baer}}$. As pointed out in that reference, however,
  the reader should be warned that elsewhere in the
  literature, e.g.\ Refs.~\cite{Choi:2005uz,Choi:2007ka} by Choi
  et al., $\alpha$ is defined such that 
  $\alpha_{\text{Choi}}=\frac{16\pi^2}{\log(M_P/m_{3/2})}
  \frac{1}{\alpha_{\text{Baer}}}$. The coupling ratio as a function of
  $\alpha$ will of course be a different function of $\alpha_{\text{Choi}}$.}
\begin{equation}
  M_1:M_2:M_3 \simeq (\alpha+3.3):(2\alpha+1):(6\alpha-9). \label{eq:MM_Ratio}
\end{equation}
This distinct ratio of couplings is a hallmark sign of this model, and
suggests that measuring the weak-scale gaugino masses could enable
one to determine $\alpha$.

In pure anomaly mediation, there is a well-known problem of tachyonic
slepton masses, which in the standard mAMSB scenario is solved by
adding an ad hoc common term to all the sfermion masses. In this
model, for moderate values of $\alpha$ the tachyonic slepton problem
is solved by the gravity mediation contributions, and this term
does not need to be included.

%
%
%
%
%
%

%
%
%
%
%

\section{MMAMSB model point} {\label{sec:Model_point}}
In order to carry out a fit in the MMAMSB parameter space, we first
need to select a point in this space to serve as our model. For a given
point in parameter space, then, we must compute the spectrum and check
that various constraints are satisfied and desirable features are
present.

We use the \texttt{ISASUGRA} spectrum generator that is packaged with
\texttt{ISAJET} \cite{Paige:2003mg} to compute the RGE running of the
parameters and to compute the low-energy spectrum and decays. 
We then use a number of codes to check various constraints: the \texttt{IsaTools} package 
provided with \texttt{ISAJET} and the programs \texttt{SuperIso Relic} \cite{Arbey:2009gu} 
and \texttt{micrOMEGAs} \cite{Belanger:2010gh,Belanger:2004yn,Belanger:2001fz}.
Each of these codes computes the relic density and various precision low energy and
flavor constraints, and \texttt{IsaTools} and \texttt{micrOMEGAs} also compute
dark matter direct detection cross sections. We also check that the model
has not been ruled out by Higgs searches using \texttt{HiggsBounds}
\cite{Bechtle:2011sb}.


We consider scenarios with $1\lesssim\alpha\lesssim10$, that have roughly similar SUSY breaking contributions from each sector and are close to the original KKLT construction value of $\alpha=5$.

In order for a fit using standard LHC observables, especially the kinematic
endpoints, to be feasible, the decay chain
\begin{equation}
  \sq\to q\neu2\to q\ell^{\pm}\sell_R^{\pm}\to\ell^+\ell^-\neu1
  \label{decaychain}
\end{equation}
must be present with a sufficiently large branching ratio.
This means that the $\sq\to\neu2$ and $\neu2\to\sell_R$ decays must have
sizeable branching ratios and sufficiently large mass splittings to give 
rise to hard enough jets and leptons to pass experimental cuts.
In addition, the squark masses must be sufficiently
low to give reasonable squark production cross sections at the LHC, but not so low that
they are clearly ruled out by early LHC searches.

For simplicity, we choose a common value of the modular weights $n_i$ for all of the
sfermion and Higgs fields. In order to satisfy the branching ratio and mass splitting
requirements just mentioned,
we find that in this case $n=1/2$ is the only valid choice. There is some freedom
in choosing $\tan\beta$ and $\text{sign}(\mu)$, so we adopt the SPS1a \cite{Allanach:2002nj} values
$\tan\beta=10$ and $\text{sign}(\mu)=+1$.

Performing a scan over the remaining parameters $\alpha$ and \mgrv, we find that the 
strongest constraint on finding a viable model is the measurement by the WMAP experiment
\cite{Komatsu:2008hk} of the dark matter relic density.
The computed relic density takes on a value within the observed error bars in 
only narrow strips of the $\alpha-\mgrv$ plane\footnote{
  For a given parameter point, the calculated relic density can vary significantly depending
  on which code is used to compute it and on which spectrum calculator is used. Because the
  relic density depends on the spectrum, and thus the model parameters, this difference can be undone by a slight change in the parameters and thus
  does not affect our results.}.
One such region is an almost-vertical strip at $\alpha\simeq 4.8$, with \mgrv ranging from 
\unit{15}{\TeV} upwards, where a mostly bino though somewhat mixed \neu1
undergoes efficient annihilation into $h\,A$, $b\,\bar{b}$, $W^+\,W^-$ and $Z\,H$. From this region we select
the benchmark point with $\mgrv=\unit{21}{\TeV}$. The model parameters 
and the weak-scale gaugino mass parameters are summarized in Tab.~\ref{tab:InputParam}.

\begin{table}[t!] 
  \begin{center}
    \begin{tabular}{|cc|cc|} \hline
      Parameter             & Value            & Parameter & Value \\ \hline \hline
      $\alpha$              & 4.8              & $M_1$     & 460.6 \\  
      $m_{3/2}$              & 21$\times10^{3}$  & $M_2$     & 556.0 \\ 
      $\tan\beta$           & 10               & $M_3$     & 976.7 \\  
      $\text{sign}(\mu)$    & +1               &           &       \\  
      $n$                   & 0.5              &           &       \\ \hline 
    \end{tabular}
    \caption{MMAMSB benchmark point. All masses in GeV. \label{tab:InputParam}}
  \end{center}
\end{table}

This model satisfies all the observational constraints
implemented in the codes mentioned above, though the spin-independent
direct detection cross section is right on the edge of the XENON100 \cite{Aprile:2011hi} 
limit. The value of $\alpha$ is close to the value
preferred by the KKLT string scenario ($\alpha_{\text{KKLT}}=5$), and is far enough from
the gravity-mediated and anomaly-mediated limits to give distinct phenomenology. In
particular, the gaugino mass ratio (at the weak scale) in this model is
$M_1:M_2:M_3\simeq 1:1.2:2.2$, which is quite distinct from the ratio
$1:2:6$ expected from gravity mediation, and $3.3:1:9$ expected from anomaly mediation.
The model also has the branching ratios and mass differences mentioned above
that are required for the standard LHC observables.

The full spectrum of the model is given Tab.~\ref{tab:ModelMass}, and is plotted in
Fig.~\ref{fig:specCompare} alongside the SPS1a spectrum for
comparison. Benchmark planes and lines for further LHC studies of
MMAMSB models have 
been proposed in Ref.~\cite{AbdusSalam:2011fc}
\begin{table}[t!] 
\begin{center}
\begin{tabular}{|cc|cc|cc|} \hline
  Particle      & Mass  & Particle & Mass  & Particle   & Mass   \\ \hline \hline
  $\sd_L$       & 939.3 & $\sell_L$ & 535.9 & $\neu{1}$  & 439.8  \\  
  $\sd_R$       & 907.5 & $\sell_R$ & 478.5 & $\neu{2}$  & 498.8  \\ 
  $\su_L$       & 935.7 & $\stau_1$ & 469.4 & $\neu{3}$  & 529.2  \\  
  $\su_R$       & 908.3 & $\stau_2$ & 533.8 & $\neu{4}$  & 611.7  \\  
  $\tilde{b}_1$ & 837.6 &           &       & $\chpm{1}$ & 487.6  \\  
  $\tilde{b}_2$ & 899.0 &           &       & $\chpm{2}$ & 608.7  \\
  $\st_1$       & 613.2 &           &       & $\sg$      & 1021.8 \\
  $\st_2$       & 894.2	&           &       &            &        \\\hline 
\end{tabular}
\caption{MMAMSB benchmark model. All masses in GeV. \label{tab:ModelMass}}
\end{center}
\end{table}

\begin{figure}
  \includegraphics[width=0.48\textwidth]{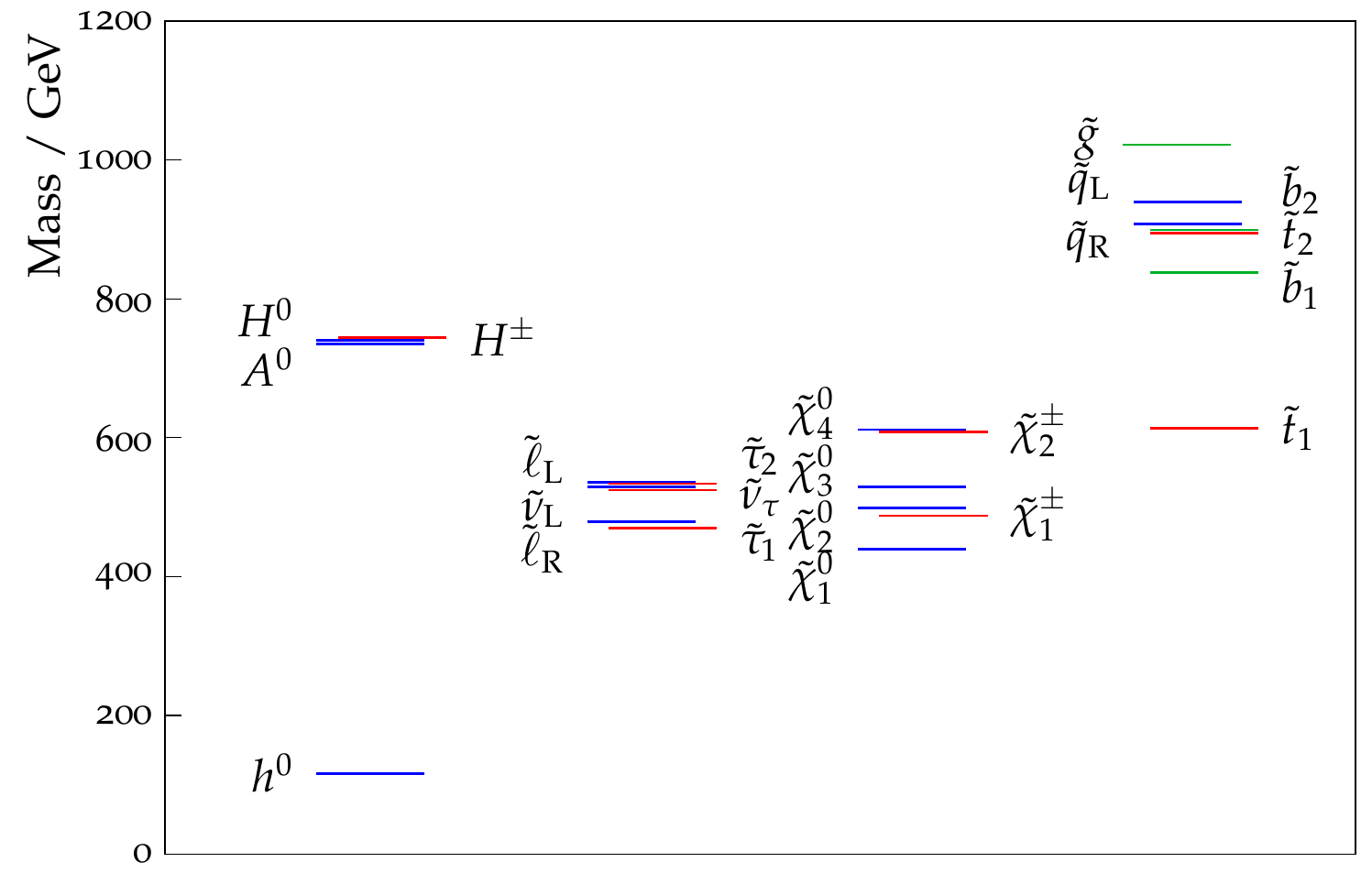} \hfill
  \includegraphics[width=0.48\textwidth]{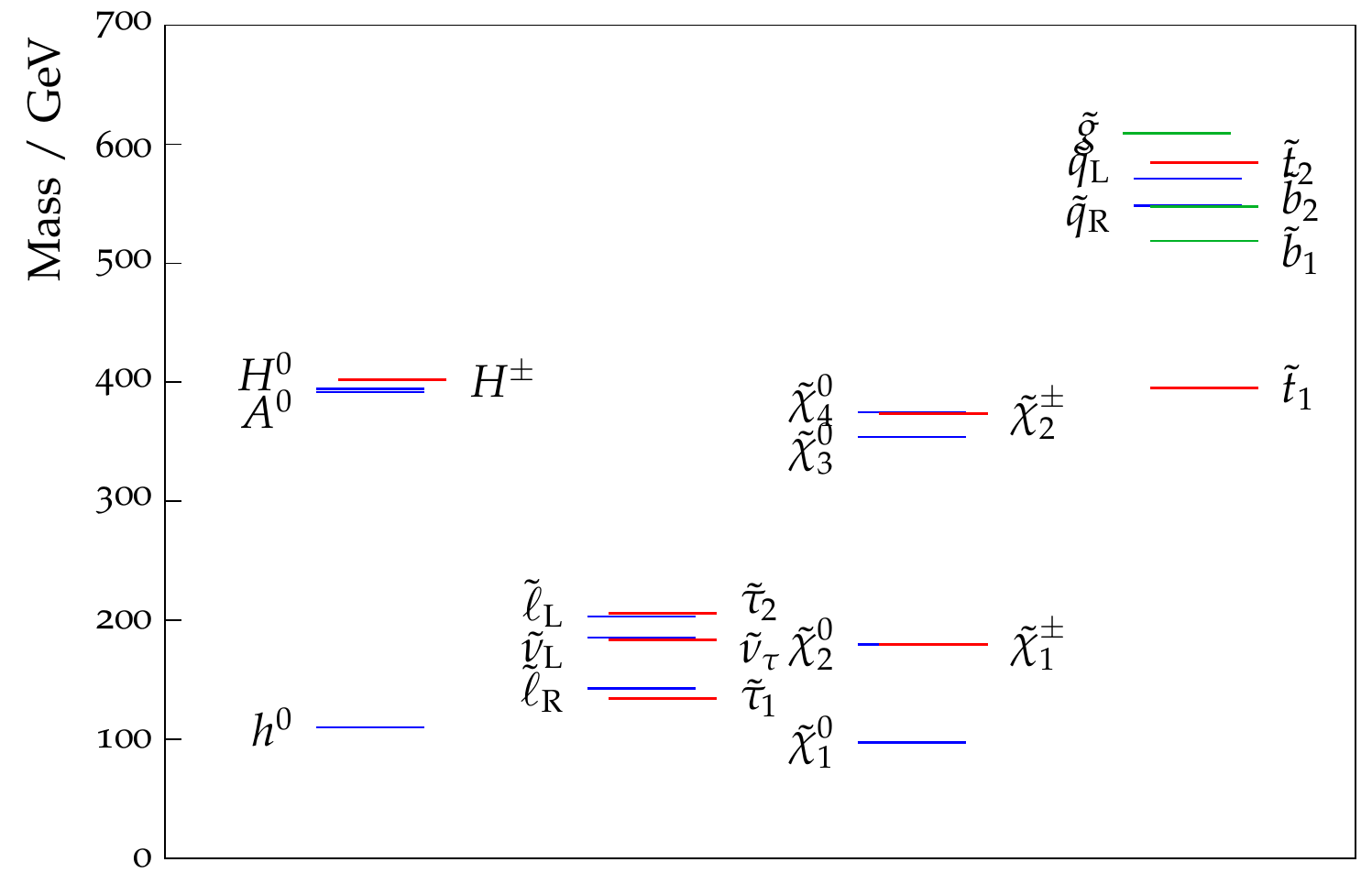}
  \caption{The spectrum of our MMAMSB model (on the left) compared to the spectrum
    of SPS1a (on the right). Note the difference in scales on the vertical axis.\label{fig:specCompare}}
\end{figure}

\section{Fit procedure} {\label{sec:Fit_proc}}
In order to perform the fits we use the \ftno program \cite{Bechtle:2004pc}.
\ftno first requires an input file, in which the user specifies the set of observables and provides their
(perhaps hypothetical) measured values and uncertainties. Here the user also specifies the supersymmetric model
and the high or low scale parameters to be fitted along with their starting values.
\ftno then efficiently samples the parameter space, finds the best fit point
and maps out confidence regions ({\it i.e.}, contours of the likelihood function)
using Markov chain Monte Carlo.

It is important to emphasize that in the fits we perform here, we only use LHC observables
as inputs. While it is possible with \ftno to also include in the fits the various low-energy, flavor, and 
astrophysical observables that were used in the previous section to pick a benchmark model, 
we focus instead on the power of the LHC alone to constrain and measure the model parameters.

In order to compute the likelihood function at each point in parameter space
visited by the Markov chain, \ftno first calls an external code 
to calculate the mass spectrum, decay widths, and branching ratios.
We use the \texttt{ISASUGRA} spectrum calculator, since it includes the MMAMSB scenario.

The predicted values of the chosen observables are then computed internally by \ftno or
by an external code, and compared with the measured values to compute the likelihood.
Originally, \ftno included among its LHC observables a large set of kinematical
edges, as well as some branching ratio observables. Crucial to our 
analysis will be the inclusion of event rates, which,
as mentioned earlier, were implemented as
observables in \ftno in Ref.~\cite{Dreiner:2010gv}. 

In this implementation,
\ftno first uses the decay table computed by the spectrum calculator to
determine the branching ratios of the squark- and gluino-initiated
decay chains that could contribute to the given channel. Each branching
ratio is then multiplied by the relevant squark and/or gluino production 
cross section. This cross section is interpolated from a grid over
\msq and \msg of NLO cross sections that have been previously computed
using the code \texttt{Prospino} \cite{Beenakker:1996ed,Beenakker:1996ch,Beenakker:1997ut}.
It remains to compute the acceptance, i.e.\ the fraction of events which
pass the analysis cuts for the given channel. As is described in 
detail in Ref.~\cite{Dreiner:2010gv}, in this implementation the acceptance
is computed by combining analytical formulae for the momentum distributions
of the decay products of at-rest squarks and gluinos with numerical
estimates (which are also interpolated from pre-computed grids) 
of the effect of boosting the system into the lab frame.

It was shown in Ref.~\cite{Dreiner:2010gv}
that this fast technique for estimating the acceptance agrees to within
5\% with acceptances determined 
using Monte Carlo event generation. To verify that this agreement holds
within the parameter space of the MMAMSB scenario, we computed
acceptances across a grid in MMAMSB parameter space using 
\texttt{Herwig++} \cite{Gieseke:2011na,Bahr:2008pv} to generate events
and \texttt{Rivet} \cite{Buckley:2010ar}, including the anti-$k_t$ jet finder 
\cite{Cacciari:2005hq,Cacciari:2008gp}, to apply the cuts. 
We then performed some of our MMAMSB fits using acceptances
interpolated from these grids alongside the fits using the rates
code implemented in \ftno, and found very good agreement.




An extremely important ingredient in the calculation of the likelihood is the
estimate of the uncertainty on each observable. For the well-studied SPS1a 
model, which is the basis of the \ftno fits performed in Refs.~\cite{Dreiner:2010gv,Bechtle:2009ty},
the uncertainties for most observables can be obtained from the thorough study carried out
in Ref.~\cite{Weiglein:2004hn}. For the fit we perform here, we must extrapolate
these uncertainties from SPS1a to our MMAMSB model. To do this, we must take into
account the relative number of events from SPS1a and our model in each signal channel.
In our MMAMSB model, the squarks and gluino have higher masses than they do in SPS1a,
so the production cross sections are lower, leading to fewer signal events
and thus larger uncertainties on most observables. In some cases,
however, an increase in branching ratios between models can more than make up for
the decrease in cross section. For example, the branching ratio of the decay 
$\neu{4}\to \sell^{\pm}_R \ell^{\mp}$ is ten times larger in our model than in
SPS1a, leading to a much more pronounced di-lepton edge due to $\neu{4}$ decays.

To verify that the scaling of uncertainties we performed is reasonable, we
looked at one of the observables, the $m_{\ell\ell}$ endpoint (for a more detailed
description of each observable, see the next two subsections) in detail.
We generated events for SPS1a and our MMAMSB model, including energy smearing 
to account for detector effects. We then fit
a smeared step function to the data in order to extract the endpoint and 
determine its uncertainty. The uncertainties we obtained in this way roughly match
the SPS1a uncertainties given in Ref.~\cite{Weiglein:2004hn} and the uncertainties
we extrapolated from this for our MMAMSB model, verifying the reasonableness of
our extrapolations. For illustration, in Fig.~\ref{fig:mllfits} we show the $m_{\ell\ell}$
distribution, with statistical error bars,
 for each model for a particular choice of center-of-mass energy and integrated 
luminosity, with the fitted function superimposed over the data.

\begin{figure}
  \centering
  \includegraphics[width=0.45\textwidth]{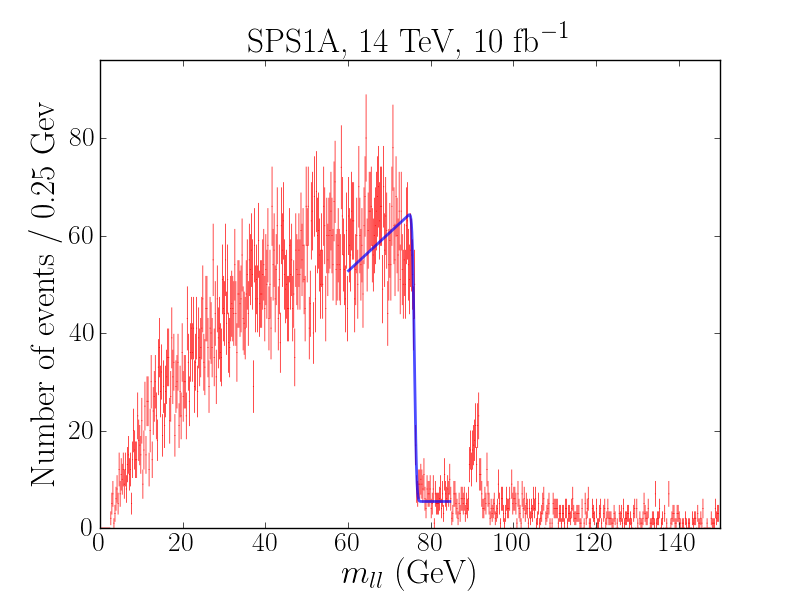}
  \includegraphics[width=0.45\textwidth]{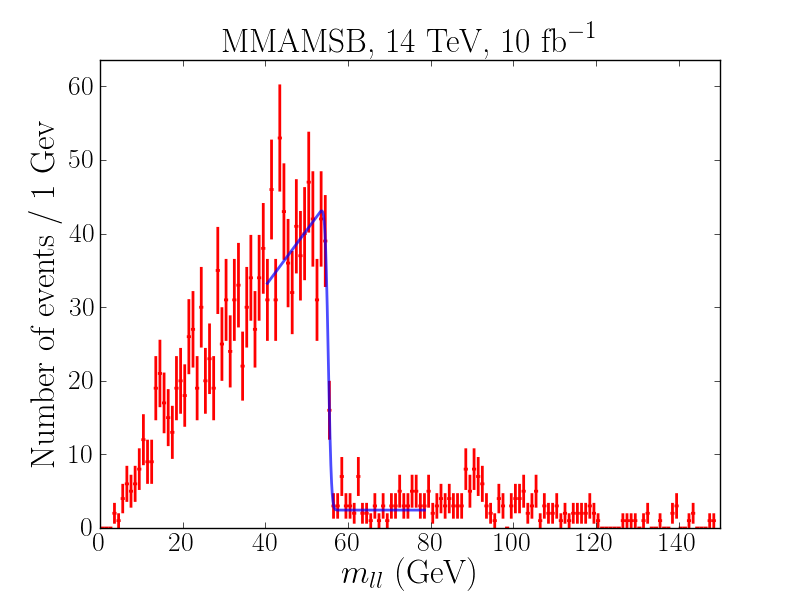}
  \caption{The $m_{\ell\ell}$ distribution for SPS1a (left) and the MMAMSB model (right)
    for \unit{10}{\invfb} of integrated luminosity at the LHC with $\sqrt{s}=\unit{14}{\TeV}$.
    The error bars show the statistical error only. Superimposed is the fit function used
    to extract the endpoint value.\label{fig:mllfits}}
\end{figure}


\subsection{Kinematic edges}
Throughout this paper, we compare the impact of various sets of observables on 
the fits. The first two sets below do not include rates, and are mostly edges of 
kinematical distributions.
The basic set we call Group I and includes standard kinematical edges
built from the ``golden'' decay chain described earlier.
Group II contains additional quantities built from this decay chain
as well as some observables sensitive to the properties of third
generation sparticles.
The full list of Group I and II observables is as follows:
\begin{itemize}
\item List of observables in Group I (these are all defined in Ref.~\cite{Gjelsten:2004ki}).
  \begin{itemize}
  \item $m_{\ell\ell}^{\mathrm{max}}$, the dilepton invariant mass edge.
  \item $m_{q\ell\ell}^{\mathrm{max}}$, the jet dilepton invariant mass edge.
  \item $m_{q\ell}^{\mathrm{low}}$, the jet-lepton low invariant mass edge.
  \item $m_{q\ell}^{\mathrm{high}}$, the jet-lepton high invariant mass edge.
  \end{itemize}
\item List of observables in Group II.
  \begin{itemize}
  \item $m_{q\ell \ell}^{\mathrm{thr}}$, the jet-dilepton threshold invariant mass edge \cite{Gjelsten:2004ki}. 
  \item $m^{T2}_{\sq}$, the squark stransverse mass \cite{Lester:1999tx,Barr:2003rg}.
  \item $m_{\tau \tau}^{\mathrm{max}}$, the di-tau invariant mass edge \cite{Weiglein:2004hn,Bechtle:2009ty}.
  \item $m_{tb}^{w}$, the weighted top-bottom invariant mass edge \cite{Bechtle:2009ty}. 
  \item $\Delta m_{\sg \neu1}$, the mass difference between the gluino and the LSP \cite{Weiglein:2004hn,Bechtle:2009ty}.  
  \item $m^{\mathrm{max}}_{(\neu4)\ell\ell}$, the dilepton invariant mass edge from the decay of a $\neu{4}$ \cite{Weiglein:2004hn,Bechtle:2009ty}.
  \item $r_{\tilde{\ell} \stau}^{\mathrm{BR}}$, the ratio of selectron (smuon) to stau mediated $\neu{2}$ decays \cite{Bechtle:2009ty}.
  \end{itemize}
\end{itemize}

In Tab.~\ref{tab:ObsErr} we provide, for each observable, its nominal value in our MMAMSB model and its
uncertainty, determined by scaling from SPS1a as described above, for each luminosity and center-of-mass energy scenario. 
The last two columns give additional contributions to the uncertainty
from the lepton energy scale and jet energy scale \cite{Bechtle:2009ty}.

\begin{table}[t!]  
  \begin{center}
    \begin{tabular}{cc|cccccc|} \cline{3-8}
      & & \multicolumn{6}{c|}{Uncertainty} \\ \hline
      \multicolumn{1}{|c|}{\multirow{2}{*}{Observable}} & \multicolumn{1}{c|}{Nominal} 
      & \unit{10}{\invfb} & \unit{1}{\invfb} & \unit{10}{\invfb} & \unit{100}{\invfb} & \multirow{2}{*}{LES} & \multirow{2}{*}{JES} \\
      \multicolumn{1}{|c|}{}                            & \multicolumn{1}{c|}{value} 
      & \unit{7}{\TeV}  & \unit{14}{\TeV}  & \unit{14}{\TeV}   & \unit{14}{\TeV}    &                      & \\\hline \hline
      \multicolumn{1}{|c|}{\bf{Group I}}                   &         &	     &		&		&		&	& \\ 
      \multicolumn{1}{|c|}{$m_{\ell \ell}^{\mathrm{max}}$}    & 55.45   & 6.01     & 4.25       & 1.34        & 0.43         & 0.05  &   -    \\ 
      \multicolumn{1}{|c|}{$m_{q\ell \ell}^{\mathrm{max}}$}      & 373.4   & 70.2      & 49.6       & 15.7        & 4.96         & -     &   3.7    \\ 
      \multicolumn{1}{|c|}{$m_{q \ell}^{\mathrm{low}}$}        & 223.3   & 38.0       & 26.8       & 8.5         & 4.40        & -     &   2.2    \\  
      \multicolumn{1}{|c|}{$m_{q \ell}^{\mathrm{high}}$}         & 311.9   & 26.0       & 18.4       & 5.8         & 4.70         & -     &   3.1    \\ \hline \hline
      \multicolumn{1}{|c|}{\bf{Group II}}                &         &            &            &		&		&	& \\ 
      \multicolumn{1}{|c|}{$m_{q\ell \ell}^{\mathrm{thr}}$}       & 145.5  & -           & -   & 29.6        & 9.37          & -     &   1.5    \\ 
      \multicolumn{1}{|c|}{$m^{T2}_{\sq}$}                    & 662.0  & -           & -    & 28.2        & 8.91        & -     &   7.0    \\ 
      \multicolumn{1}{|c|}{$m_{\tau \tau}^{\mathrm{max}}$}       & 58.94   & -           & -   & 15.9        & 5.04         & -     &   0.6    \\  
      \multicolumn{1}{|c|}{$m_{tb}^{w}$}                     & 494.1   & -           & -   & 43.0        & 13.6        & -     &   4.9    \\ 
      \multicolumn{1}{|c|}{$\Delta m_{\sg \neu1}$}            & 582.0    & -          & -  & 48.5        & 15.3        & -     &   5.8    \\ 
      \multicolumn{1}{|c|}{$m^{\mathrm{max}}_{(\neu4)\ell\ell}$}   & 168.6    & -          & -  & 9.96        & 3.15        & 0.17     &   -    \\ 
      \multicolumn{1}{|c|}{$r_{\tilde{\ell} \stau}^{\mathrm{BR}}$} & 0.457  & -           & -    & 0.0114      & 0.0036     & -     &   -    \\ \hline
    \end{tabular}
    \caption{LHC observables for the MMAMSB benchmark point, Tab.~\ref{tab:InputParam}. The masses and branching ratios 
      have been calculated with \texttt{ISASUGRA} \cite{Paige:2003mg}. The uncertainty estimates on the observables are 
      based on \cite{Weiglein:2004hn,Bechtle:2009ty} and have been rescaled as described in the main text. All dimensionful quantities are given in GeV. \label{tab:ObsErr}}
  \end{center}
\end{table}

\subsection{Rates}
In addition to the Group I and II observables listed in the previous subsection,
we also consider the two rate observables discussed previously. Here, we give the description
of these observables, including the full list of cuts.
\begin{itemize}
\item List of rate observables. (These observables are also defined in Ref.~\cite{Dreiner:2010gv}.)
  \begin{itemize}
  \item $R_{jj\slashed{E}_T}$, the event rate for at least two hard jets and missing transverse energy.
    \begin{itemize}
    \item $p_{T,\text{jet}} > 50$ GeV.
    \item $|\eta_{\text{jet}}| < 2.5$.
    \item $R_{\text{jet}}=0.4$ (anti-${k_t}$ jet algorithm \cite{Cacciari:2005hq,Cacciari:2008gp}).
    \item $\slashed{E}_T > 100$ GeV.
    \end{itemize}  
  \item $R_{\ell\ell jj\slashed{E}_T}$, the event rate for at least two hard jets and missing transverse energy 
    with a pair of opposite sign, same flavour leptons (1st or 2nd generation). The background from the 
    leptonic decays of $\tau$ leptons, $\chpm{1,2}$ and $W^{\pm}$ is removed by subtracting events with 
    opposite sign, different flavour lepton pairs.
    \begin{itemize}
    \item $R_{jj\slashed{E}_T}$ signal.
    \item $p_{T,\ell} > 10 {\mathrm GeV}$.
    \item $|\eta_\ell| < 2.5$.
    \item Lepton-jet isolation of $\Delta R=0.2$ (jet activity within $\Delta R=0.2 < 10$ GeV).
    \end{itemize}  
  \end{itemize}
\end{itemize}

In Tab.~\ref{tab:RateErr}, we provide the values for these observables, {\it i.e.} the event rates
for the two different types of events as predicted by a Monte Carlo simulation. We provide the
rates and their uncertainties for both 7 and \unit{14}{\TeV}.
\begin{table}[t!] 
  \begin{center}
    \begin{tabular}{|c|cc|cc|} \hline
                               &  \multicolumn{2}{c|}{\unit{7}{\TeV}}  & \multicolumn{2}{c|}{\unit{14}{\TeV}}  \\
      Observable               & Value (fb)  & Uncertainty  	& Value	(fb)   & Uncertainty \\ \hline\hline
      $R_{jj\slashed{E}_T}$        & 113    & 23            & 2780     & 556          \\  
      $R_{\ell\ell jj\slashed{E}_T}$ & 11.8   & 3.5           & 245      & 49           \\ \hline 
    \end{tabular}
    \caption{LHC event rates for the MMAMSB benchmark point, Tab.~\ref{tab:InputParam}. The event rate includes 
      the NLO squark and gluino production cross section \cite{Beenakker:1996ch,Beenakker:1997ut}, the branching ratios of the decays and the expected particle acceptances. 
      The acceptances were tested with full parton shower and hadronisation using 
      \texttt{Herwig++} \cite{Gieseke:2011na,Bahr:2008pv}, \texttt{Rivet} \cite{Buckley:2010ar} and the anti-$k_t$ 
      jet finder \cite{Cacciari:2005hq,Cacciari:2008gp}. \label{tab:RateErr}}
  \end{center}
\end{table}
\section{Fit results}  {\label{sec:Fit_res}}

In this section we present the main results of our paper and show that including event rates in fits can significantly increase their effectiveness to measure and exclude different SUSY breaking scenarios. We begin by fitting to our own MMAMSB benchmark scenario and show that even with early data, the model can be constrained and some parameters measured accurately. With larger data sets, we should be able to measure the whole of the parameter space with high accuracy ($\lesssim5\%$). 

We then try to fit other SUSY breaking scenarios to our benchmark point from MMAMSB. We start with the CMSSM and show that if we only include the Group I mass edges, a good fit can be achieved unless we have very large data sets (\ifb{100} at \unit{14}{\TeV}). However, if we include event rates in the fit, we see that even with a small amount of data (\ifb{10} at \unit{7}{\TeV}), we can exclude the CMSSM. We complete the same task for the mAMSB and see very similar results. With Group I observables and early data we cannot conclusively exclude the scenario. However, as soon as we add event rates, the model can no longer fit the data.

\subsection{MMAMSB}

In this section we discuss the results of fitting the MMAMSB model to observables derived from our benchmark scenario. Fits are performed 
for \ifb{10} at \tev7 and \ifb1, \ifb{10}, and \ifb{100} at \tev{14}. 
In addition, we also show the effect of 
adding rates (see Tab.~\ref{tab:RateErr}) to the fits, and including the more complicated set of Group II observables (see Tab.~\ref{tab:ObsErr}). 
We demonstrate that rates significantly improve the accuracy of the fits, especially when using early LHC data with limited statistics.

Let us first consider the fits that can be performed with \ifb{10} at \tev7 and \ifb1 at \tev{14}. With this amount of data
and our benchmark scenario, we do not expect to have sufficient statistics to be able to measure any of the observables given in 
Group II. Therefore, we perform the fits using the mass edges in Group I, both with and without the rates. 
In the left-hand column of Figs.~\ref{fig:7TeV_Comp} and \ref{fig:1fb_Comp}, we see that when only using 
the Group I edges, the model is essentially unconstrained across the whole parameter space and an effective 
fit cannot be performed. 

\begin{figure}\centering
  \includegraphics[width=0.85\textwidth]{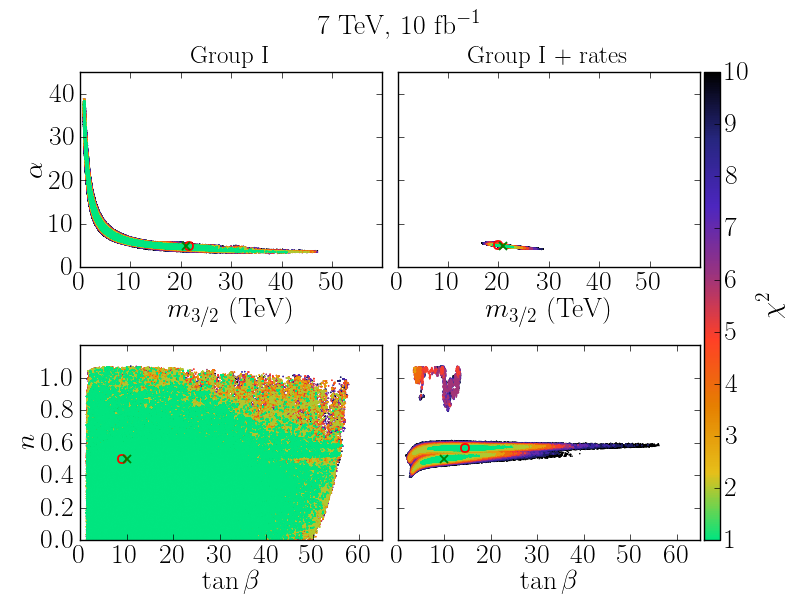}
  \caption{\label{fig:7TeV_Comp} 1-$\sigma$ (68.3\%) (black), 2-$\sigma$ (95.4\%) (grey), 
    and 3-$\sigma$ (99.7\%) (light grey) 
    two-dimensional confidence regions for \tev7, \ifb{10}, using Group I observables,
    without (left column) and with (right column) rates. The green \textquoteleft X' represents
    the input point and the red circle the best-fit point.}
\end{figure}

\begin{figure}\centering
  \includegraphics[width=0.85\textwidth]{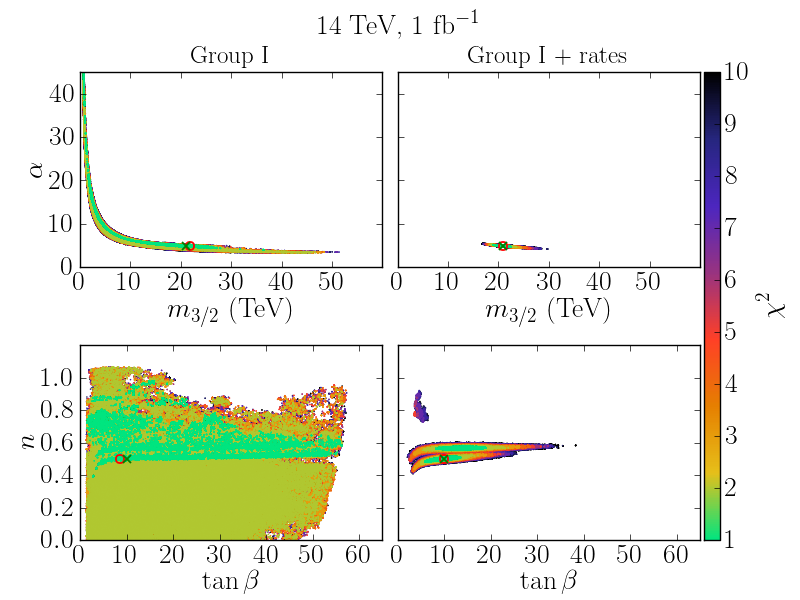}
  \caption{\label{fig:1fb_Comp} Like Figure~\ref{fig:7TeV_Comp} except for
    \tev{14}, \ifb1.}
\end{figure}

As soon as we add rates to the fit, however, the situation improves remarkably. At our benchmark point we can now constrain 
$m_{3/2}$ to $\sim$ 15\% and $\alpha$ to $\sim$ 10\% as can be seen in
the right-hand column of Figs.~\ref{fig:7TeV_Comp} and \ref{fig:1fb_Comp}, or in the zoomed-in version
of these regions shown in Fig.~\ref{fig:Zoom_Comp}. 
The fact that the rates improve the constraints 
on $m_{3/2}$ and $\alpha$ is to be expected if we examine the form of the soft breaking terms in the MMAMSB model, 
Eqs.~(\ref{eq:Gauge_MM})-(\ref{eq:Scalar_MM}). We see that the soft breaking masses are all proportional to the product
$\alpha m_{3/2}$. Therefore, the rate observables, which are sensitive to the overall mass scale of any new states, constrain 
this combination effectively. The modular weight, $n$, is less well constrained---at \tev7 
it is only determined within $\sim$ 20\%, as can be seen in 
Fig.~\ref{fig:Zoom_Comp}---due to the fact that it has a sub-leading dependence in the soft breaking terms. 
This is still an improvement, however, over the fit without rates, where $n$ could not be constrained at all. Finally, $\tan\beta$ 
has the worst constraints of all the parameters in the fit; even with rates included, most values are allowed. 
As we show later, a measurement of $\tan\beta$ requires observables that are sensitive to third-generation 
particles and none of the observables in Group I measure these. 

\begin{figure}\centering
  \includegraphics[width=0.85\textwidth]{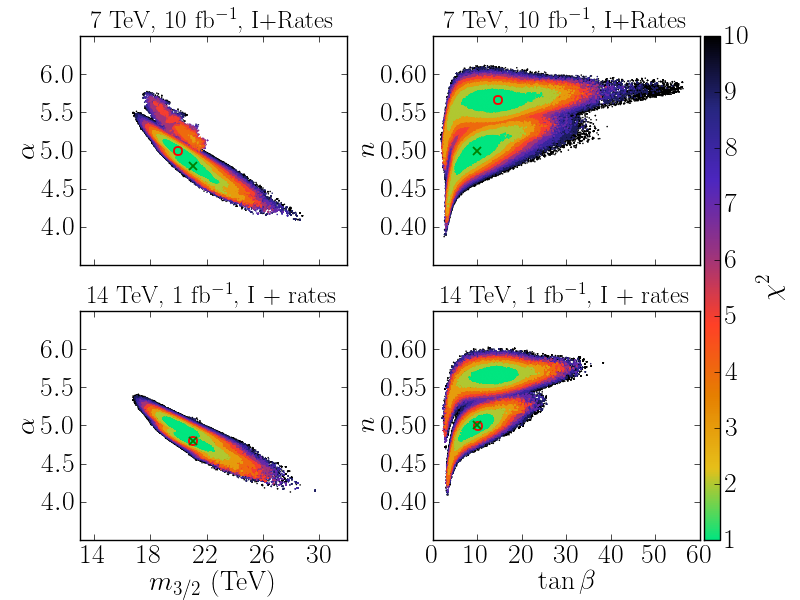}
  \caption{\label{fig:Zoom_Comp} A zoomed-in view of the confidence regions for 
    \tev7, \ifb{10} and \tev{14}, \ifb1.
    These are the same fits as in Fig.~\ref{fig:7TeV_Comp} and Fig.~\ref{fig:1fb_Comp}.}
\end{figure}

If we move to the fits with \ifb{10} at \tev{14}, we see in  Fig.~\ref{fig:10fb_Comp1} that even an accurate determination of all the 
Group I mass edges can still not constrain the MMAMSB model
in any parameter. Once we add the information from rates, we again see a good fit for $m_{3/2}$, $\alpha$ and $n$, and
even $\tan\beta$ is constrained ($\tan\beta=10^{+8.5}_{-3.5}$). The reason that $\tan\beta$ can be constrained with 
the rate observables is that the combination of $R_{jj\slashed{E}_T}$ and $R_{\ell\ell jj\slashed{E}_T}$ 
acts as an observable for the branching ratio of the decay $\neu{2}\to\sell^{\pm}\ell^{\mp}$. This branching ratio is sensitive 
to $\tan\beta$ since it competes with the branching ratio of the decay $\neu{2}\to\stau^{\pm}\tau^{\mp}$, which increases substantially as 
$\tan\beta$ becomes large.

\begin{figure}\centering
  \includegraphics[width=0.85\textwidth]{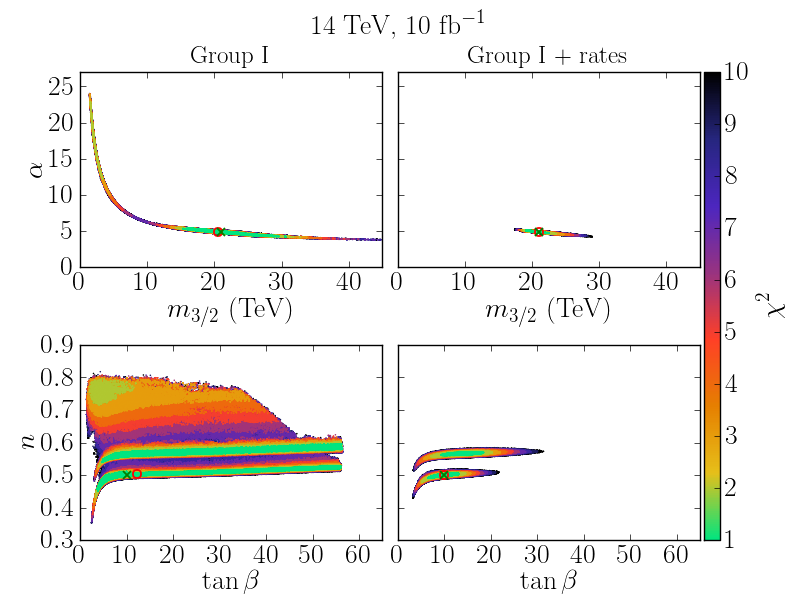}
  \caption{\label{fig:10fb_Comp1} Confidence regions for \tev{14}, \ifb{10}, only including Group I observables,
    with (left column) and without (right column) rates.}
\end{figure}

Adding the Group II set of observables with a reasonable statistical error becomes possible with \ifb{10} at \tev{14}. 
We can now compare the fit using the Group I and II observables alone with that of using just the Group I observables along with rates. We 
see in Fig.~\ref{fig:10fb_Comp2} that in general the fit is greatly improved, especially for $\tan\beta$ and $n$. The constraining power 
for $\tan\beta$ mainly comes from the ratio $r_{\tilde{\ell} \stau}^{\mathrm{BR}}$ and the mass edge $m_{\tau\tau}$. These observables directly measure 
the contribution of third generation superpartners, which is most sensitive to $\tan\beta$. The measurement of the mass scale $m_{3/2}$ is 
not improved significantly, however, as this was already well constrained by the rate information. In the Group II observables, the squark stransverse 
mass ($m_{\sq}^{T2}$) and the jet-dilepton threshold ($m_{qll}^{thr}$) now perform this task. 

\begin{figure}\centering
  \includegraphics[width=0.85\textwidth]{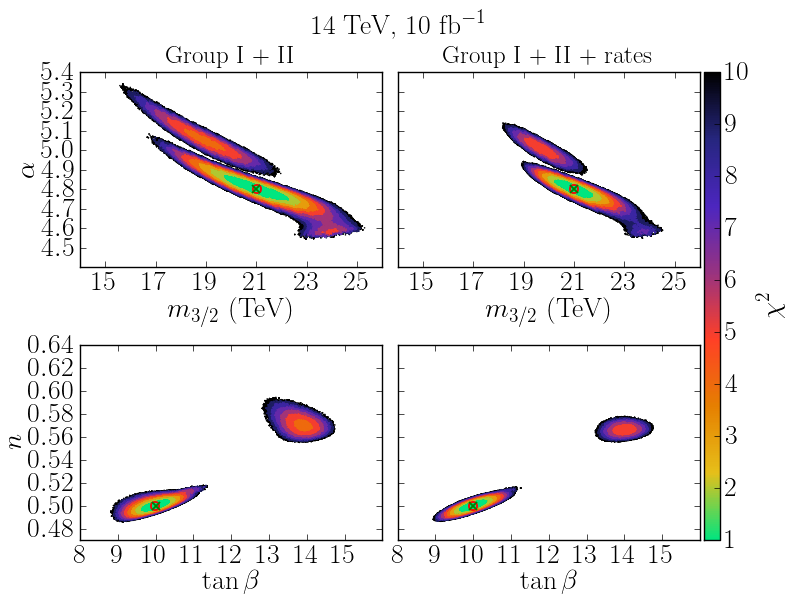}
  \caption{\label{fig:10fb_Comp2} Like Fig.~\ref{fig:10fb_Comp1}, but now also including Group II observables.}
\end{figure}

Despite the extra observables that constrain the mass scale in Group II, if we add rates to the fit, the measurement of both $m_{3/2}$ 
and $\alpha$ improves substantially, as is clear from Fig.~\ref{fig:10fb_Comp2}. As stated before, this is due to the rates being particularly sensitive 
to the overall mass scale. The measurement of $m_{3/2}$ is improved by almost a factor of 2.

An interesting feature of the fits that now becomes much more clear is the double minimum seen in the $\tan\beta$--$n$ plane. 
The double minimum is due to the fact that the functional forms for 
$m_{qll}^{\mathrm{max}}$, $m_{ql}^{\mathrm{low}}$ and $m_{ql}^{\mathrm{high}}$ depend on the relative mass ratios of the particles in the cascade 
decay \cite{Allanach:2000kt,Gjelsten:2004ki}. For example, in the correct minimum we have a mass ordering
$2m^2_{\sell} > m^2_{\neu{1}}+m^2_{\neu{2}} > 2m_{\neu{1}}m_{\neu{2}}$, which leads to the end-points given in Tab.~\ref{tab:ObsErr}. 
However, in the second minimum, the mass ordering switches to $m^2_{\neu{1}}+m^2_{\neu{2}} > 2m_{\neu{1}}m_{\neu{2}} > 2m^2_{\sell}$, 
where the end-points have a different functional form. Thus, a different set of masses can lead to the same end-points being 
measured at the LHC. The problem of end-point mimics was previously been discussed in Refs.~\cite{Gjelsten:2005sv,Gjelsten:2006as} 
and possible solutions proposed were to measure the \neu1 mass accurately at a linear collider \cite{Martyn:2004ew,Conley:2010jk} or to measure the whole 
invariant mass distribution as the shape would be different for the mimic distribution. Unfortunately, the rates are unable to resolve the end-point mimics as they are only weakly sensitive to the masses of the electroweak particles further down the decay chain. The rates are most sensitive to the first generation squarks and gluino masses and in both minima, these are very similar. 

With \ifb{100} of data, the Group I mass edges are all measured with high accuracy. Consequently, these mass edges alone 
can constrain all of the parameters apart from $\tan\beta$ to some extent, as can be seen in Fig.~\ref{fig:100fb_Comp1}. Adding the rates 
to the fit, though, improves the measurements significantly. We see a large improvement in the determination of $m_{3/2}$ 
(the width of the 1$\sigma$ confidence region drops from \tev{10} to \tev4) and $\alpha$, while $\tan\beta$ is becoming constrained.

\begin{figure}\centering
  \includegraphics[width=0.85\textwidth]{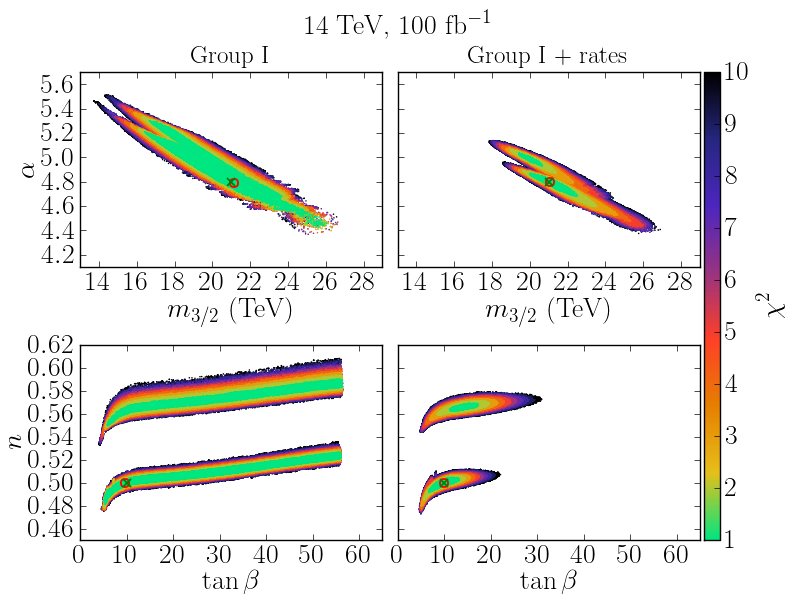}
  \caption{\label{fig:100fb_Comp1} Like Fig.~\ref{fig:10fb_Comp1}, but using \ifb{100}.}
\end{figure}

If we include Group II observables to the above fits, as shown in Fig.~\ref{fig:100fb_Comp2}, we now achieve a precision determination of better 
than 5\% on all parameters. In fact, $\alpha$, which parameterises the ratio of gauge couplings [see Eq.~(\ref{eq:MM_Ratio})] is measured 
to better than 1\%. In addition, the double minimum displayed in other fits is no longer present and we now have no mass ordering 
ambiguity. The observable that breaks the ambiguity is $m_{\tau\tau}^{\mathrm{max}}=\gev{32.0}$ (measured \gev{58.9}) which is 5.3~$\sigma$ away. 
This is tensioned against $m_{\ell\ell}^{\mathrm{max}}$ which is the most accurately measured observable. If we want to increase 
$m_{\tau\tau}^{\mathrm{max}}$ for this point, we would also increase $m_{\ell\ell}^{\mathrm{max}}$, leading to an even worse fit. 

\begin{figure}\centering
  \includegraphics[width=0.85\textwidth]{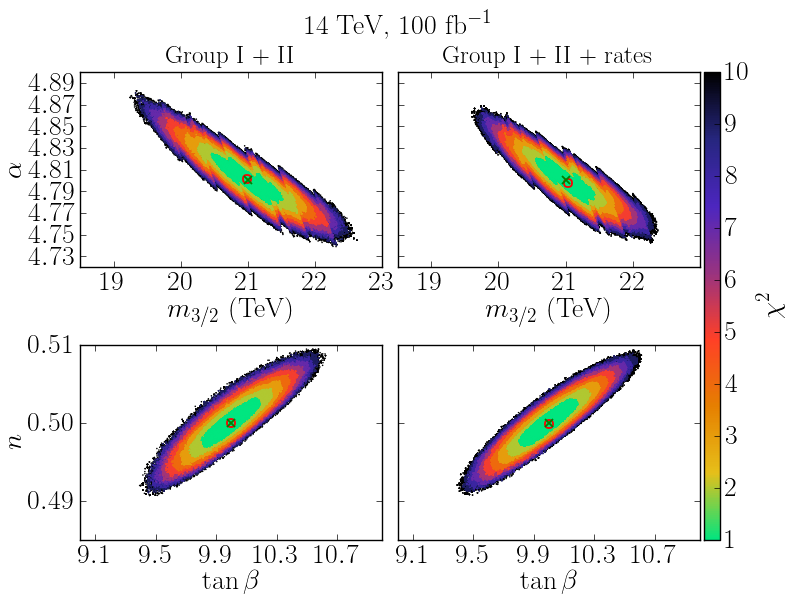}
  \caption{\label{fig:100fb_Comp2} Like Fig.~\ref{fig:10fb_Comp2}, but using \ifb{100}.}
\end{figure}


In the high accuracy fits using \ifb{100} (Figs.~\ref{fig:100fb_Comp1} and \ref{fig:100fb_Comp2}), two interesting correlations 
become apparent. First, in the $m_{3/2}$--$\alpha$ plane, there is a negative correlation between the two parameters. This correlation can be 
easily understood when we inspect the form of the soft breaking terms in the model, Eqs.~(\ref{eq:Gauge_MM})-(\ref{eq:Scalar_MM}). 
All of the soft masses are $\propto \alpha m_{3/2}$. Therefore to keep the particle masses at a set value, as $m_{3/2}$ increases, 
$\alpha$ must decrease and vice versa.

The other clear correlation is a positive one between the parameters $\tan\beta$ and $n$. This correlation is due to the trilinear 
coupling $A_{ijk}$ which enters in the \stau mixing matrix. The mixing term in the \stau sector is
\begin{equation}
  m_{\stau}^{mix} = m_{\tau} (A_{\tau}-\mu\tan\beta)\,,
\end{equation}
and from Eq.~(\ref{eq:Trilinear_MM}) we can see that $A_{\tau}\propto n$. Therefore, in order to keep the mixing in the \stau sector constant, 
if $n$ $(A_{ijk})$ increases $\tan\beta$ must also increase to compensate. The opposite would be true of the \st sector but the 
observables $r_{\tilde{\ell} \stau}^{\mathrm{BR}}$ and $m_{\tau \tau}^{\mathrm{max}}$ are far more accurately measured than the observable
$m_{tb}^{w}$. Thus the fit is dominated by the constraint in the \stau sector.

The results of our MMAMSB fit are collected in Tab.~\ref{tab:MMAMSB_Fit}.

\begin{table}[t!] \renewcommand{\arraystretch}{1.3}
  \begin{center}
    \begin{tabular}{|c|cccc|} \hline
                               &  $\alpha$            & $m_{3/2}$~(TeV)    & $\tan\beta$         & n      \\
      MMAMSB                   & 4.8                  & 21                & 10                 & 0.5    \\ \hline \hline
      \bf{\tev7 and \ifb{10}}    &                       &                   &                    &        \\
      I                        & $4.8_{-1.4}^{+33.5}$    & $22_{-21}^{+19}$     & $9_{-8}^{+48}$       & $0.5_{-0.5}^{+0.5}$    \\  
      I + rates                & $4.99_{-0.42}^{+0.15}$  & $20.0_{-1.0}^{+2.9}$  & $15_{-10}^{+10}$     & $0.56_{-0.10}^{+0.02}$  \\ \hline \hline
      \bf{\tev{14} and \ifb1}    &                       &                   &                    &       \\
      I                        & $4.8_{-0.8}^{+41.0}$    & $22_{-21}^{+15}$     & $9_{-7}^{+48}$       & $0.5_{-0.1}^{+0.5}$     \\  
      I + rates                & $4.80_{-0.13}^{+0.31}$  & $21.0_{-2.1}^{+1.5}$  & $10_{-4}^{+9}$       & $0.50_{-0.02}^{+0.08}$  \\ \hline \hline
      \bf{\tev{14} and \ifb{10}}   &                       &                    &                    &       \\
      I                        & $4.8_{-0.6}^{+0.5}$     & $21_{-5}^{+10}$      & $12_{-9}^{+44}$       & $0.50_{-0.05}^{+0.09}$  \\  
      I + rates                & $4.80_{-0.12}^{+0.26}$   & $21.0_{-1.9}^{+1.5}$  & $10_{-3}^{+9}$       & $0.50_{-0.01}^{+0.07}$  \\ 
      I + II                   & $4.80_{-0.05}^{+0.07}$   & $21.0_{-1.3}^{+1.2}$  & $10.0_{-0.3}^{+0.4}$  & $0.500_{-0.004}^{+0.005}$ \\  
      I + II + rates           & $4.80_{-0.04}^{+0.04}$   & $21.0_{-0.7}^{+0.7}$  & $10.0_{-0.3}^{+0.4}$  & $0.500_{-0.004}^{+0.005}$ \\ \hline \hline
      \bf{\tev{14} and \ifb{100}}  &                        &                   &                    &       \\
      I                        & $4.8_{-0.4}^{+0.3}$      & $21_{-4}^{+5}$      & $10_{-4}^{+47}$       & $0.50_{-0.02}^{+0.09}$     \\  
      I + rates                & $4.80_{-0.12}^{+0.24}$    & $21.0_{-1.6}^{+1.5}$ & $10_{-3}^{+7}$        & $0.500_{-0.008}^{+0.069}$  \\ 
      I + II                   & $4.801_{-0.023}^{+0.024}$ & $21.0_{-0.5}^{+0.5}$ & $9.99_{-0.19}^{+0.19}$  & $0.500_{-0.003}^{+0.003}$  \\  
      I + II + rates           & $4.798_{-0.019}^{+0.023}$ & $21.0_{-0.5}^{+0.4}$ & $10.00_{-0.19}^{+0.19}$  & $0.500_{-0.003}^{+0.003}$ \\ \hline 
    \end{tabular}
    \caption{Fits to MMAMSB parameters for our chosen benchmark point. Fits are done with various sets 
      of observable groups (I and II) and errors (Tab.~\ref{tab:ObsErr}). Fits are also done with and 
      without the rates observables (Tab.~\ref{tab:RateErr}). \label{tab:MMAMSB_Fit}}
  \end{center}
\end{table}

\subsection{CMSSM}

\begin{table}[t!] \renewcommand{\arraystretch}{1.3}
  \begin{center}
    \begin{tabular}{|c|cccc|c|} \hline
      CMSSM                   &  $m_0$~(GeV)        & $m_{1/2}$~(GeV)     & $\tan\beta$     & $A_0$~(GeV)        		& $\chi^2$/d.o.f.  \\ \hline \hline
      \bf{\tev7 and \ifb{10}}   &                    &                    &                 &                    		&                          \\
      I                       & $36_{-21}^{+189}$     & $210_{-58}^{+12}$    & $5_{-3}^{+40}$     & $405_{-1056}^{+1256}$     &   0.12/0           \\  
      I + rates               & $78$               & $413$              & $7.8$           & $649$              			& 216/2          \\ \hline \hline
      \bf{\tev{14} and \ifb1}   &                    &                    &                 &                    	          &                  \\
      I                       & $35_{-12}^{+59}$      & $208_{-21}^{+10}$    & $4_{-1.0}^{+29}$   & $409_{-1038}^{+1237}$ 	& 0.23/0      \\  
      I + rates               & $69$               & $379$              & $7.6$           & $580$              			& 334/2                 \\ \hline \hline
      \bf{\tev{14} and \ifb{10}}  &                    &                    &                 &                    	         &                   \\
      I                       & $35.3_{-4.8}^{+47.8}$ & $208.4_{-10.1}^{+3.2}$ & $5_{-2}^{+27}$    & $373_{-742}^{+801}$   	&  2.1/0           \\  
      I + rates               & $59$               & $331$              & $9.4$           & $538$              			& 1643/2           \\ 
      I + II                  & $39$               & $210$              & $8.0$           & $364$              			& 122/7           \\  
      I + II + rates          & $57$               & $328$              & $6.5$           & $531$              			& 1806/9              \\ \hline \hline
      \bf{\tev{14} and \ifb{100}} &                    &                    &                 &                  		  &                               \\
      I                       & $33.6_{-2.1}^{+2.5}$  & $207.3_{-2.4}^{+2.1}$ & $4.7_{-1.2}^{+2.2}$& $365_{-105}^{+112}$   	& 11.8/0          \\  
      I + rates               & $51$               & $319$              & $8.0$           & $542$             			 & 2533/2            \\ 
      I + II                  & $38$               & $203$              & $8.1$           & $354$             			 & 907/7            \\  
      I + II + rates          & $173$              & $311$              & $5.8$           & $502$             			 & 4043/9            \\ \hline
    \end{tabular}
    \caption{Best fit points for the CMSSM and the minimum $\chi^2$ for that point and associated set of observables. We only include the 1-$\sigma$ 
      environment when the best fit point is not excluded at the 99.9\% confidence level. We see that rates are extremely effective at ruling 
      out the CMSSM.  \label{tab:CMSSM_Fit}}
  \end{center}
\end{table}

In the previous section we have shown that we can accurately reconstruct our MMAMSB benchmark point at the LHC, even with relatively 
small amounts of data (\ifb{10} at \tev{14}). Here we would like to determine, however, if it is also possible to rule out other SUSY breaking
scenarios. Possibly the most widely studied scenario is the CMSSM and thus it is natural to ask what happens if we attempt to fit
this scenario to our benchmark point. In particular we would like to use the $\chi^2$ of the best-fit point to determine if the
CMSSM can be ruled out. 

We begin by performing the fit for \ifb{10} at \tev7 using only the Group I observables and without the information from rates.\footnote{For 
  these fits the number of measurements is equal to the number of free parameters. Therefore, 
  a perfect fit to the data ($\chi^2=0$) should in principle always be possible, but this is not the case, as is shown by 
  the fits at \tev{14} with \ifb{10} and \ifb{100} where $\chi^2=2.1$ and $\chi^2=11.8$ respectively. There are a couple of reasons for this.
  One is that the  Group I observables only measure the wino, bino and first generation sfermion masses. Thus, they only have weak sensitivity 
  to $\tan\beta$ and $A_0$ and the fit is effectively to only two free parameters. A second reason is that 
  many other constraints are implicitly present in the fit but not formally included in the number of measurements. For 
  example, we require a neutral LSP so that we have a dark matter candidate, consistent electroweak symmetry breaking, and 
  no tachyonic degrees of freedom. All of these constrain the areas of 
  parameter space that we are able to search to satisfy the mass-edge conditions.} 
The results of this fit are shown in Tab.~\ref{tab:CMSSM_Fit}. 
We see that the fit has a viable minimum with a $\chi^2$ of just 0.12. In addition, the fit begins to 
constrain the parameter $m_{1/2}$ ($160 < m_{1/2} < 220$) while the other parameters can vary quite freely, as is illustrated in Fig.~\ref{fig:MSUGRA_Fit}. 
As we move to \ifb1 at \tev{14} but keep the same observable set, constraints begin to appear on $m_0$ ($0 < m_{0} < 100$) and we still 
have a minimum where all the observables are well fitted ($\chi^2=0.23$). 

\begin{figure}\centering
  \includegraphics[width=0.85\textwidth]{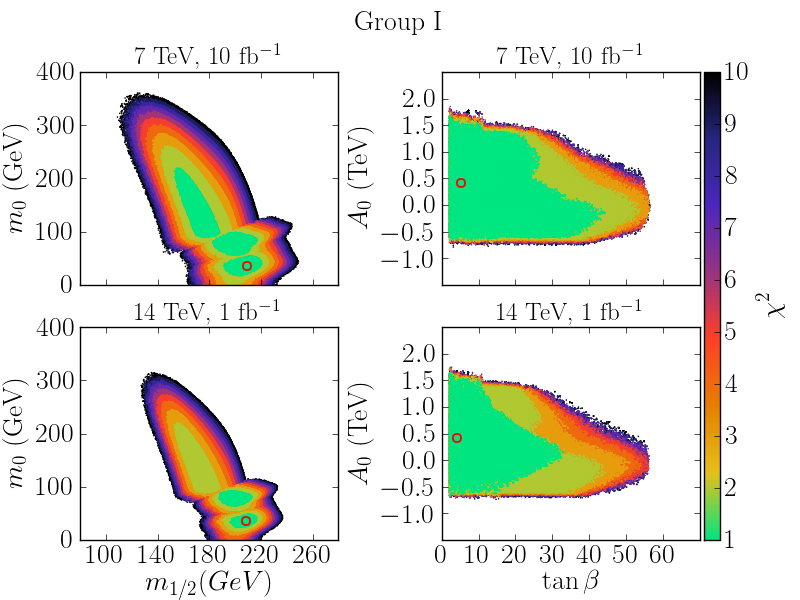}
  \caption{\label{fig:MSUGRA_Fit} Plots showing the fit of the CMSSM to our MMAMSB benchmark point using only Group I observables and no 
    rate information. The fits in the upper row are for \tev7, \ifb{10} while the bottom row shows the fits for \tev{14}, \ifb1.}
\end{figure}

If we add the rate information to the fit, however, the result changes spectacularly and the best fit has $\chi^2=216$ which means 
that the point is completely excluded by the data. If we examine the individual measurements at this best fit point more closely, we see 
that the exclusion is dominated by two measurements. The jet-lepton high invariant mass edge, whose true value is $m_{ql}^{\mathrm{high}}=\gev{652}$, 
has a measured value of \gev{311.9} at the best-fit point, a full 13~$\sigma$ away. The event rate at the best fit point, $R_{jj\met}=\unit{231}{\femtobarn}$ 
(measured = \unit{113}{\femtobarn}), is 5.3~$\sigma$ away. Essentially, the CMSSM model cannot replicate the more compressed mass spectrum with the relatively large 
masses for the gluino and squarks. The edge observables (especially the difference between the squark and slepton mass) are trying 
to pull the mass scale down, while the rate observables try to pull the mass scale up. The tension between these two observables 
leads to the exclusion.

We can also examine the best-fit point found using the Group I mass edges alone to see what rates are predicted for these points. 
We find that for the best-fit point with \ifb{10} at \tev7, $R_{jj\met}=14809$ (measured = 113) and is 652~$\sigma$ away! Incidentally,
this best fit point is already easily ruled out by current searches 
\cite{ATLAS-CONF-2011-090,jetsPlusMissingEPS,Chatrchyan:2011qs,Chatrchyan:2011ek}. 

Moving to the fits performed at \tev{14} with \ifb{10}, we see that using the Group I observables alone, $\chi^2=2.1$, and thus the fit still 
has a viable minimum. If we add the Group II observables, however, the best-fit point now has a $\chi^2$/d.o.f.= 122/7
and is thus excluded. The exclusion is dominated by the stransverse mass observable, which at the best-fit point is
$m_{\sq}^{T2}=\gev{444}$, 8.9~$\sigma$ away from the measured value $m_{\sq}^{T2}=\gev{699}$. The stranverse mass is effective as its rough dependence is 
$m_{\sq}^{T2}\propto\sqrt{m^2_{\sq}-2m^2_{\neu{1}}}$ and thus with heavy squarks, it is essentially measuring the squark mass. This leads to a similar tension 
as in the fits with rates: the stransverse mass tries to pull the mass scale of the fit up while the mass edges try to bring the scale down. 

Even though the exclusion due to the stransverse mass seems convincing, we should compare this to the exclusion that would be achieved 
by using the rate observables. At the best-fit point, the inclusive rate has the predicted value $R_{jj\met}=\unit{125}{\picobarn}$, which is a huge 
218~$\sigma$ away from the value $R_{jj\met}=\unit{2.78}{\picobarn}$ that would be measured.
Thus we can see that the rates are far more effective observables for determining the overall mass scale of a 
model than the stransverse mass is. This can also be seen by comparing the fit done with \ifb{10} at \tev7 with only the Group 
I observables and including rates with the fit done with \ifb{10} at \tev{14} with Group I and II observables but no rate information. 
From Tab.~\ref{tab:CMSSM_Fit}, we see that even though the fit with rates is done with fewer observables in total and far less data, 
with a total $\chi^2=216$ it more convincingly rules out the CMSSM model than the Group II fit (total $\chi^2=122$) that would take at least a few years 
longer.

\subsection{mAMSB}

\begin{table}[t!] \renewcommand{\arraystretch}{1.3}
  \begin{center}
    \begin{tabular}{|c|ccc|c|} \hline
      mAMSB                   &  $m_0$~(\GeV)        & $m_{3/2}$~(\TeV)       & $\tan\beta$       &  $\chi^2$/d.o.f. \\ \hline \hline
      \bf{\tev7 and \ifb{10}} &                      &                      &                             &                 \\
      I                       & $127_{-21}^{+14}$    & $15.2_{-1.8}^{+1.2}$ & $21_{-19}^{+2}$           &   3.8/1          \\  
      I + rates               & $317$                & $32$                 & $33$                  &   238/3            \\ \hline \hline
      \bf{\tev{14} and \ifb1}    &                      &                      &                    &                 \\
      I                       & $127_{-16}^{+10}$    & $15.2_{-1.4}^{+0.8}$    & $21_{-10}^{+2}$    &   7.6/1  \\  
      I + rates               & $316$                & $32$                 & $4.7$                 &   397/3         \\ \hline \hline
      \bf{\tev{14} and \ifb{10}}   &                      &                      &                   &                 \\
      I                       & $124$		   & $15$   			& $21$      	    &   72/1             \\  
      I + rates               & $316$                 & $32$                & $25$                   &   3084/3             \\ 
      I + II                  & $116$                & $14$               & $16$                    &   330/8            \\  
      I + II + rates          & $316$                 & $32$                & $9$                   &   4135/10             \\ \hline \hline
      \bf{\tev{14} and \ifb{100}}  &                      &                      &                   &                 \\
      I                       & $126$ 		   & $15$   		& $21$  		    &   275/1             \\  
      I + rates               & $292$                 & $30$                & $11$                   &   4591/3             \\ 
      I + II                  & $100$                 & $13$                & $16$                    &   1886/8             \\  
      I + II + rates          & $292$                & $30$                & $9$                    &   13678/10             \\ \hline
    \end{tabular}
    \caption{Best fit points for the mAMSB and the minimum $\chi^2$ for that point and associated set of observables. We only include the 1-$\sigma$ 
      environment when the best fit point is not excluded at the 99.9\% confidence level. We see that rates are extremely effective at ruling 
      out the mAMSB. \label{tab:mAMSB_Fit}}
  \end{center}
\end{table}

In the previous section we have shown that the MMAMSB can be convincingly distinguished from the CMSSM 
even with early LHC data (\ifb{10} at 7~TeV). We now do a similar analysis witha different SUSY breaking scenario,
mAMSB, to see if the LHC can perform the same task and also separate this model. The mAMSB scenario nearly corresponds
to pure anomaly mediation, i.e. the $\alpha\to 0$ limit of the MMAMSB. As mentioned earlier, however, pure anomaly
mediation leads to tachyonic slepton masses. In minimal AMSB (mAMSB), this problem is solved with the addition 
of a constant contribution $m_0$ to the scalar masses at the GUT scale. Since $\alpha = 0$ means that the $n$-dependence
in the model also disappears, mAMSB has only three parameters, $m_0$, $m_{3/2}$, and $\tan\beta$.

We begin by performing the fit with only the Group I observables at \ifb{10} at 7~TeV. The results of this fit are shown in the upper
row of Fig.~\ref{fig:mAMSB_Fit}. The fit has a viable minimum, but a $\chi^2$/d.o.f. = 3.8/1. Even with 
just these four mass edge observables and early data, the model already suffers from a little tension. In addition, 
both $m_0$ ($100 < m_0 < 150$~GeV) and $m_{3/2}$ ($13.4 < m_{3/2} < 16.4$~TeV) are already relatively constrained. 
As we move to \ifb{1} at 14~TeV, the tension in the fit increases and the minimum 
now has a $\chi^2$/d.o.f. = 7.6/1. This corresponds to a p-value of less than 0.01 and shows the model is already highly 
disfavoured. The reason why the Group I observables alone can exclude the model is that the mAMSB contains a larger mass 
splitting between the gaugino masses than either MMAMSB or the CMSSM (see Sec.~\ref{sec:Model_point}).

\begin{figure}\centering
  \includegraphics[width=0.85\textwidth]{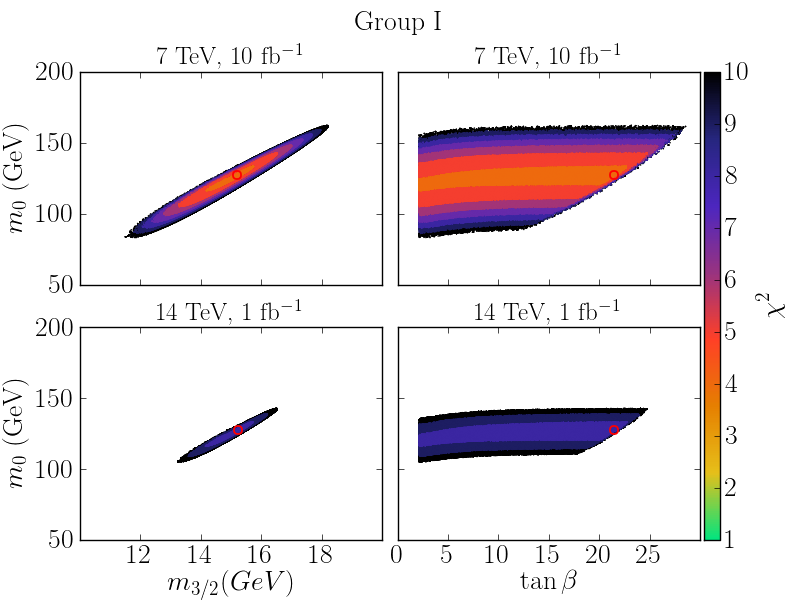}
  \caption{\label{fig:mAMSB_Fit} Plots showing the fit of mAMSB to our MMAMSB benchmark point using only Group I observables and no 
    rate information. The fits in the upper row are for \tev7, \ifb{10} while the bottom row shows the fits for \tev{14}, \ifb1.}
\end{figure}

If we include the rates in the fit, however, the exclusion of the model becomes far clearer, 
as can be seen in the lower row of Fig.~\ref{fig:mAMSB_Fit}. Including the event rates with \ifb{10} at 7~TeV 
leads to a best fit point with a minimum $\chi^2$/d.o.f. = 238/3 and the model is completely ruled out. 
The reason is the same as for the CMSSM: to fit the mass edges 
requires the mass scale in mAMSB to be low. For example, the masses of all the coloured particles in the best fit point of the Group I fit 
with \ifb{10} at 7~TeV are under 400~GeV and the event rates are far higher than they are for the model we are fitting to. 
This analysis is confirmed when we look at the individual observables. At the best fit point 
the jet-lepton high invariant mass edge has the value $m_{q\ell}^{\mathrm{high}}=653$~GeV, which is 13~$\sigma$ away from its value (312~GeV) at 
the MMAMSB benchmark point, and is trying pull the mass scale down. On the other hand, the event rate at the best fit point, 
$R_{jjE_T^{\mathrm{miss}}}=240$~fb, is 5.7~$\sigma$ away from the MMAMSB value of 113~fb and is trying
to pull the mass scale up. The tension between the two observables leads to the clear exclusion.

As for the fit to the CMSSM, we can also examine the effect of the Group II observables on the exclusion and in particular the squark stransverse mass, $M_{\sq}^{T2}$, which is sensitive to the mass scale. We find the fit with Group I and II observables but without event rates with \ifb{10} at 14~TeV has  $\chi^2/\text{d.o.f.}=330/8$ and thus excludes the model. As expected, the exclusion is dominated by $M_{\sq}^{T2}$ which is 12~$\sigma$ away from the true value at the best fit point. Once again, we have a tension between the edge observables, which are trying to pull the mass scale down, and $M_{\sq}^{T2}$, which is trying to pull the mass scale up.

In the CMSSM fit, we also compared the effectiveness of the squark stransverse mass observable with that of the event rates. Here we do the same by taking the above best fit point with just Group I and II observables and examining what the event rate would be at this point. We find that the event rate $R_{jjE_T^{\mathrm{miss}}}=\unit{539}{\picobarn}$ for the best fit point which is 953~$\sigma$ away from the value (\unit{2.78}{\picobarn}) in our MMAMSB benchmark scenario. Again, this confirms that rates are far more sensitive to the mass scale than $m_{\sq}^{T2}$ and provide a much more constraining test of the SUSY breaking scenario.

\section{Discussion} {\label{sec:Disc}}
In the MMAMSB both modulus and anomaly mediated terms can contribute at roughly equal levels to SUSY breaking. This model can have very similar phenomenology to the widely studied CMSSM and mAMSB models. A unique feature of the MMAMSB, however, is that depending on the value of the phenomenological parameter $\alpha$, the ratio of gaugino masses can be very different from other SUSY breaking scenarios.

In this paper we determined how well we can expect the MMAMSB model to be measured at the LHC and if it is possible to distinguish it from other SUSY breaking scenarios. We fit the model to various hypothetical LHC measurements to see how precisely the input parameters can be determined. Among the inputs to the fits, we included the widely studied mass edges of various SUSY cascade decay chains, as well as other kinematical observables and measurements of ratios of branching ratios that are expected to be possible at the LHC.  Special to our procedure, we also included event rate observables, to try and further constrain the model parameters. Event rates are extremely sensitive to the mass scale of the model but are conventionally difficult to include in fits due to the computationally intensive task of running a Monte Carlo event generator for each point in the fit. This is solved by using a combination of cross-section grids and analytically calculated acceptances that is vastly quicker \cite{Dreiner:2010gv}.

Using a particular MMAMSB benchmark scenario we showed that the parameters of the model will be difficult to reconstruct with early data if only mass edge measurements are used. When including the event-rate information, however, even early LHC data is sufficient to constrain the model, especially the parameters that determine the overall mass scale. With more data and more intricate observables, we showed that all the model parameters can be reconstructed with high accuracy ($\lesssim5\%$).

In addition to showing that the MMAMSB model can be reconstructed, we also demonstrated that it can be distinguished from the CMSSM and the mAMSB. Using only the basic `Group I' mass edges (see Sec.~\ref{sec:Fit_proc}), we saw that even with \ifb{100}, it is difficult to conclusively exclude the CMSSM from being able to fit our MMAMSB benchmark point. With the addition of event-rate information, however, the CMSSM interpretation is comprehensively excluded with just \ifb{10} at 7~TeV. We reached a similar conclusion when trying to fit the mAMSB to our benchmark point. Again, when we included rates, the model was easily excluded with early LHC data.

We have seen that rates play a crucial role in measuring and distinguishing different SUSY models. The models considered in this paper have few free parameters. Therefore an obvious direction for future work is to investigate more general models and see if rate information can allow the precise determination of some or all of their parameters.

\acknowledgments

The authors wish to thank Peter Wienemann for all his help in matters relating to \ftno. For particular help with including event rates into our fit we would like to thank both Ben O'Leary and Jonas Lindert. In addition, we would like to thank Hans-Peter Nilles for an introduction to the MMAMSB and related discussions. We would also like to acknowledge the help of Kiwoon Choi and Valeri L\"{o}wen.    

This work has been supported in part by the Helmholtz Alliance `Physics at the Terascale' and the DFG SFB/TR9 ``Computational Particle
Physics''. HD and JC were supported by BMBF “Verbundprojekt HEP-Theorie” under the contract 0509PDE.


\bibliography{mmamsb-refs}
\bibliographystyle{JHEP}
\end{document}